\documentclass[10pt,a4paper]{IEEEtran}
\usepackage{epsfig}
\usepackage{graphics}
\usepackage[T1]{fontenc}
\usepackage[hang,flushmargin]{footmisc}
\usepackage{flushend}
\title{
On Covert Acoustical Mesh Networks in Air
\noindent\thanks{
Originally appeared in Journal of Communications.\newline
This version is derived from \cite{hanspach.jocm} and \cite{hanspach.sich14}.
}}

\author{
Michael Hanspach
and Michael Goetz\\
\normalsize Fraunhofer FKIE, Wachtberg, Germany\\
E-Mail: \{michael.hanspach, michael.goetz\}@fkie.fraunhofer.de}

\date{}

\begin{document}

\maketitle
\pagestyle{plain}
\thispagestyle{plain}
\begin{abstract}
Covert channels can be used to circumvent system and network policies by establishing communications that have not been considered in the design of the computing system.
We construct a covert channel between different computing systems that utilizes audio modulation/demodulation to exchange data between the computing systems over the air medium.
The underlying network stack is based on a communication system that was originally designed for robust underwater communication.
We adapt the communication system to implement covert and stealthy communications by utilizing the ultrasonic frequency range.
We further demonstrate how the scenario of covert acoustical communication over the air medium can be extended to multi-hop communications and even to wireless mesh networks.
A covert acoustical mesh network can be conceived as a meshed botnet or malnet that is accessible via inaudible audio transmissions.
Different applications of covert acoustical mesh networks are presented, including the use for remote keylogging over multiple hops.
It is shown that the concept of a covert acoustical mesh network renders many conventional security concepts useless, as acoustical communications are usually not considered.
Finally, countermeasures against covert acoustical mesh networks are discussed, including the use of lowpass filtering in computing systems and a host-based intrusion detection system for analyzing audio input and output in order to detect any irregularities.
\end{abstract}


\section{Introduction}

If we want to exploit a rigorously hardened and tested type of computing system or networks of this type of computing system, we have to break new ground.
Covert channels are communication channels utilizing means for communications that have not been designed for communication at all~\cite{Lampson:1973:NCP:362375.362389}.
With a covert channel, we can circumvent system and network security policies by exploiting new, previously unregarded communication media.
In operating systems, covert channels are usually established by exploiting shared resource access between different processes, establishing a covert storage channel by encoding data in parts of the operating system that were not considered for communication at all or establishing a covert timing channel by manipulating and analyzing the timing behavior of shared resources.
In computer networks, covert storage channels can be established by utilizing normally unused parts of communication protocol headers and covert timing channels could be established over the timing behavior of network requests.

As a shared media for covert communications, not only shared computing resources or preexisting network interfaces could be used.
One can imagine a covert network channel between different computing systems, establishing a completely new network interface based on physical emanations (see also~\cite{hanspach.law}).
Alongside the established radio emanations frequently used in network communications, optical or acoustical emanations could also be used as means for communications.
Although the existing radio communication standards are mature enough for regular wireless communications, they are not used in our scenario as they are already known to the computing system's operating system and actively managed by the operating system.
For covert communications we specifically demand devices that are:

\begin{enumerate}
\item Usable as either a sending or a receiving device, i.e., they are able to output or input a physical emanation type.
\item Accessible to the sending or receiving process.
\item Not yet established as a communication device and, therefore, not subject to the system and network policies.
\item Able to support stealthy communication to prevent immediate detection of the covert channel.
\end{enumerate}

In this article, we specifically target covert communications over acoustical emanations, utilizing speakers and microphones (available in commonly available computing systems) as sending and receiving devices (see 1) that are commonly accessible (see 2) to application partitions of the operating systems that might be in need to process audio (e.g., for video conferencing or IP phone communications).
Speakers and microphones are also not established as means for communication and are not widely considered in security and network policies (see 3).
Regarding 4, stealthy communication can be implemented in acoustical networks by utilizing inaudible frequencies, i.e., the ultrasonic and near-ultrasonic frequency ranges.

We do not only target covert communication between two computing systems or two isolated application partitions on a single computing system, but we establish means for multi-hop communications between different participants.
As the underlying communication system, we utilize a specific network stack that has originally been developed for underwater communications.
Preexisting TCP/IP stacks could not reasonably be utilized for acoustical networks due to their comparably large overhead, but a loosely related addressing scheme with less overhead is utilized (see Section~\ref{covnet_sec3}).

By providing multi-hop communication, we can significantly extend the communication range of the covert channel in order to interconnect scattered devices to a full-fledged wireless mesh network.
With a covert acoustical mesh network, we can offer a whole range of \emph{covert services} to the participating computing systems, including Internet access via an IP proxy.

The remainder of this article is structured as follows.
In Section \ref{covnet_sec2}, we introduce the concept of covert networks and a scenario for covert and stealthy wireless communication between isolated computing systems.
In Section \ref{covnet_sec3}, we present the fundamental concepts of the underlying communication system that we are using for covert communications in the targeted scenario.
In Section \ref{covnet_sec4}, we describe our experimental setup for covert communications and necessary adaptions to the utilized communication system.
We construct an acoustical mesh network and describe the performance and reliability of the utilized communication system in the targeted scenario.
In Section \ref{covnet_sec5}, we present different applications for covert mesh networks.
In Section \ref{acous_counter}, we explore potential countermeasures against participation in a covert acoustical mesh network.
In Section \ref{covnet_sec7}, we discuss related work and how it differs from our work.
In Section \ref{covnet_sec8}, we conclude the article, discussing our results and giving an outlook at possible future research.

\section{Scenario}\label{covnet_sec2}

\begin{center}
\begin{figure}[!ht]
\includegraphics[width=0.5\textwidth]{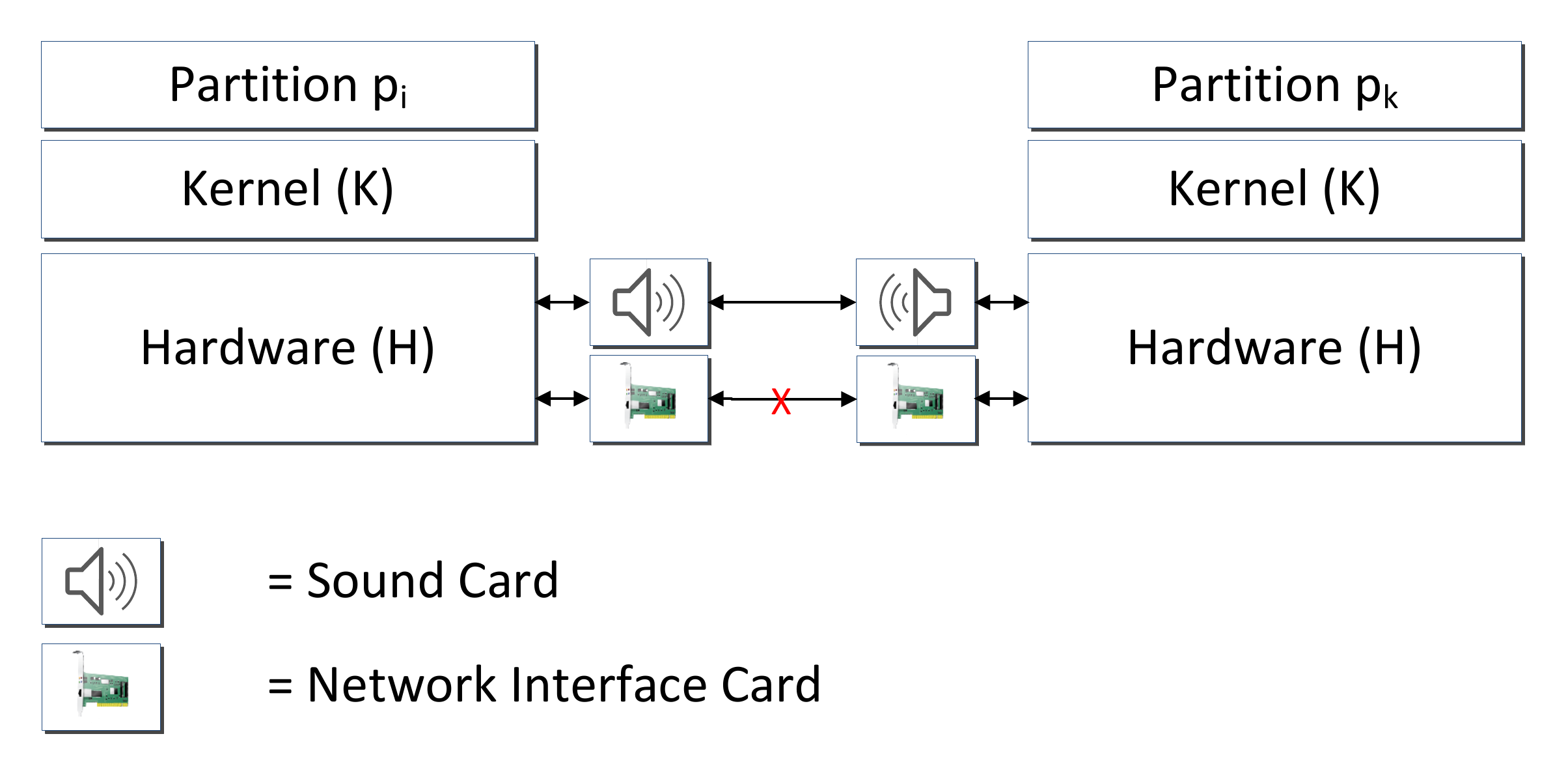}
\caption{Scenario for two computers as part of a covert network.}
\label{basic_acoustical_scenario}
\end{figure}
\end{center}

The basic scenario for covert communications between two computers is depicted in Fig.~\ref{basic_acoustical_scenario}.

Two computers that are not connected to each other via established types of network interfaces (e.g., IEEE 802.3 Ethernet~\cite{8023} or IEEE 802.11 WLAN~\cite{80211}) or that are prohibited to communicate with each other over these established types of network interfaces are, nevertheless, able to communicate with each other by using their audio input and output devices (microphones and speakers).

We assume a high-assurance setup where a component-based operating system (as described by Jaeger et al.~\cite{Jaeger:1998:SAC:319195.319229} as an operating system that consists of a small trusted computing base and individual service components) is used to define fine-grained mandatory access control policies for communications between isolated application partitions.
A component-based operating system is usually governed by a reference monitor in the kernel $K$ (as introduced by Anderson~\cite{anderson:rm}), which is an access control monitor that has always to be invoked in IPC (here: inter-partition communication) decisions.

Even in a component-based operating system that is governed by a reference monitor, acoustical communication is possible between $p_i$ and $p_k$, representing application partitions on different computing systems, as long as audio input and output devices as part of the underlying hardware $H$ are accessible to both $p_i$ and $p_k$ (the utilized terminology is also described in~\cite{hanspach.law}).
While it would be an easy solution to just disallow access to the audio hardware, this is not possible, when $p_i$ and $p_k$ are in need of audio access (e.g., for IP phone conversations).
Still there are applicable countermeasures against participation of a computing system in a covert acoustical network, as it will be presented in Section~\ref{acous_counter}.

\begin{center}
\begin{figure}[!ht]
\includegraphics[width=0.5\textwidth]{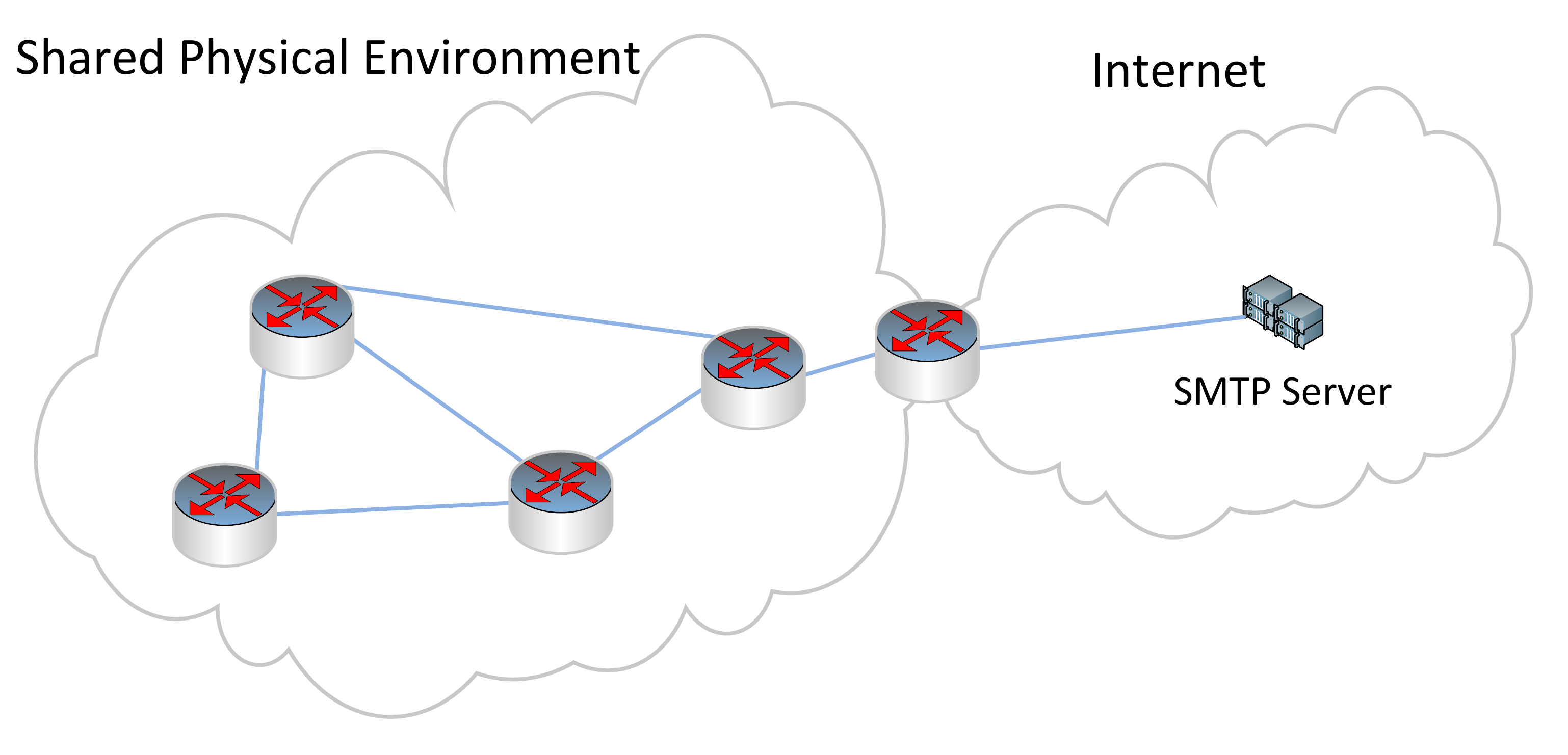}
\caption{Topology of a covert mesh network that connects a shared physical environment to an SMTP server inside the Internet.}
\label{mesh_scenario}
\end{figure}
\end{center}

We will now extend the scenario of a covert acoustical network to a covert acoustical \emph{mesh} network.
A possible topology of a covert mesh network is depicted in Fig.~\ref{mesh_scenario}.

In a covert acoustical mesh network, more than two computing systems in a shared physical environment (i.e., within the physical communication range between two connected nodes) can be connected to the mesh network and computing systems are able to communicate indirectly by following routing paths over multiple hops.
We distinguish between three types of participants in a covert mesh network:

\begin{IEEEdescription}
\item[Infected drone] 
\hfill \\
An infected computing system that offers covert services or serves as a router in the covert mesh network.
\item[Infected victim] 
\hfill \\
An infected computing system that is targeted by the attacker to secretly leak information to other participants of the covert mesh network.
\item[Attacker] 
\hfill \\
The computing system controlling the covert mesh network, and the receiver of leaked information.
\end{IEEEdescription}

Each of these participants is configured as a sender or a receiver.
In case of the infected drone, the computing system has to be configured both as a sender and a receiver, while the infected victim could just be used as a sender and the attacker could only be configured as a receiver.
All participants must have installed a compatible acoustic communication system, either by infection of a malware or actively installed (on the attacker).
The implementation details of the utilized acoustic communication system are described in the following section.

\section{Composition and Adaption of the underlying Communication System}\label{covnet_sec3}
\subsection{A Network Stack for Acoustic Communication}

Acoustical communication is seldom seen in terrestrial networks since radio offers much higher bit rates and communication ranges. 
However, acoustical communication is the method of choice in underwater networks, because electromagnetic waves are highly absorbed by sea water. 
We are, therefore, able to implement a terrestrial acoustical network on top of preliminary studies about robust underwater acoustical communication.
 
The utilized network stack is an adaption of an emulation system for underwater 
acoustical networks~\cite{wuwnetEmulator} from the Research Department for 
Underwater Acoustics and Marine Geophysics (FWG) of the Bundeswehr Technical 
Center WTD 71 in Kiel, Germany. It is split into four layers of connected applications (not to be confused with the TCP/IP stack): 

\begin{enumerate}
 \item Application Layer (APP)
 \item Network Layer (NET)
 \item Error Correction Layer (EC)
 \item Physical Link Layer (PHY)
\end{enumerate}

\begin{center}
\begin{figure}[!ht]
\includegraphics[width=0.5\textwidth]{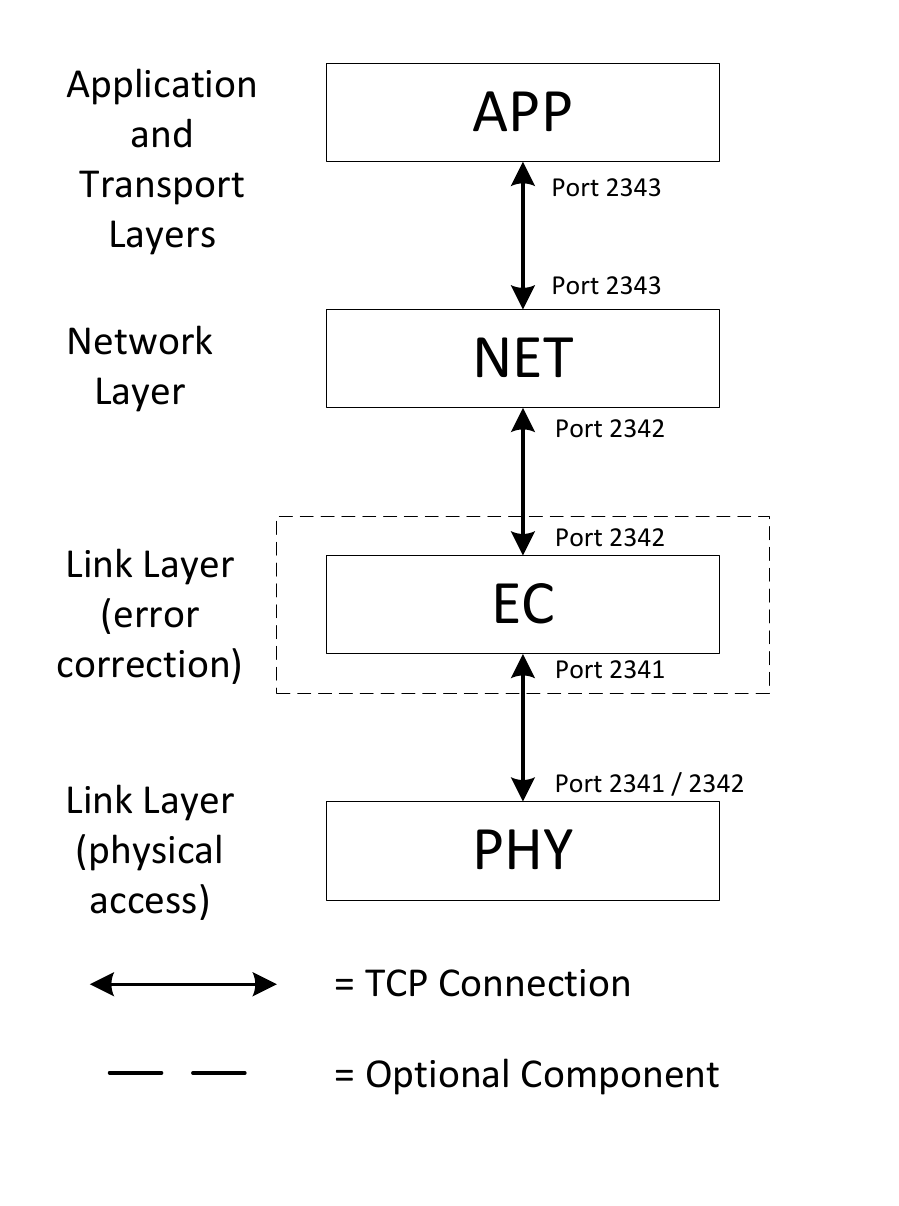}
\caption{Communication system architecture for covert acoustical mesh networks.}
\label{arch_guwal}
\end{figure}
\end{center}

All layered applications of the network stack are independent and connected via 
internal TCP connections for IPC (inter-process communication), 
as shown in Fig.~\ref{arch_guwal}. 
This structure allows to replace any of the modules, e.g., the application layer module, 
without the need to touch one of the other layers. The error correction 
layer is optional and can be left out on devices with limited processing
power or memory.

The left column in Fig.~\ref{arch_guwal} lists the layers of the TCP/IP stack, while the right column shows the associated layers/applications within the acoustic emulation system.
In the following subsections all three layers (application, network and physical link) are described in detail. 

\subsection{Application Layer}

\begin{center}
\begin{figure}[!ht]
\includegraphics[width=0.5\textwidth]{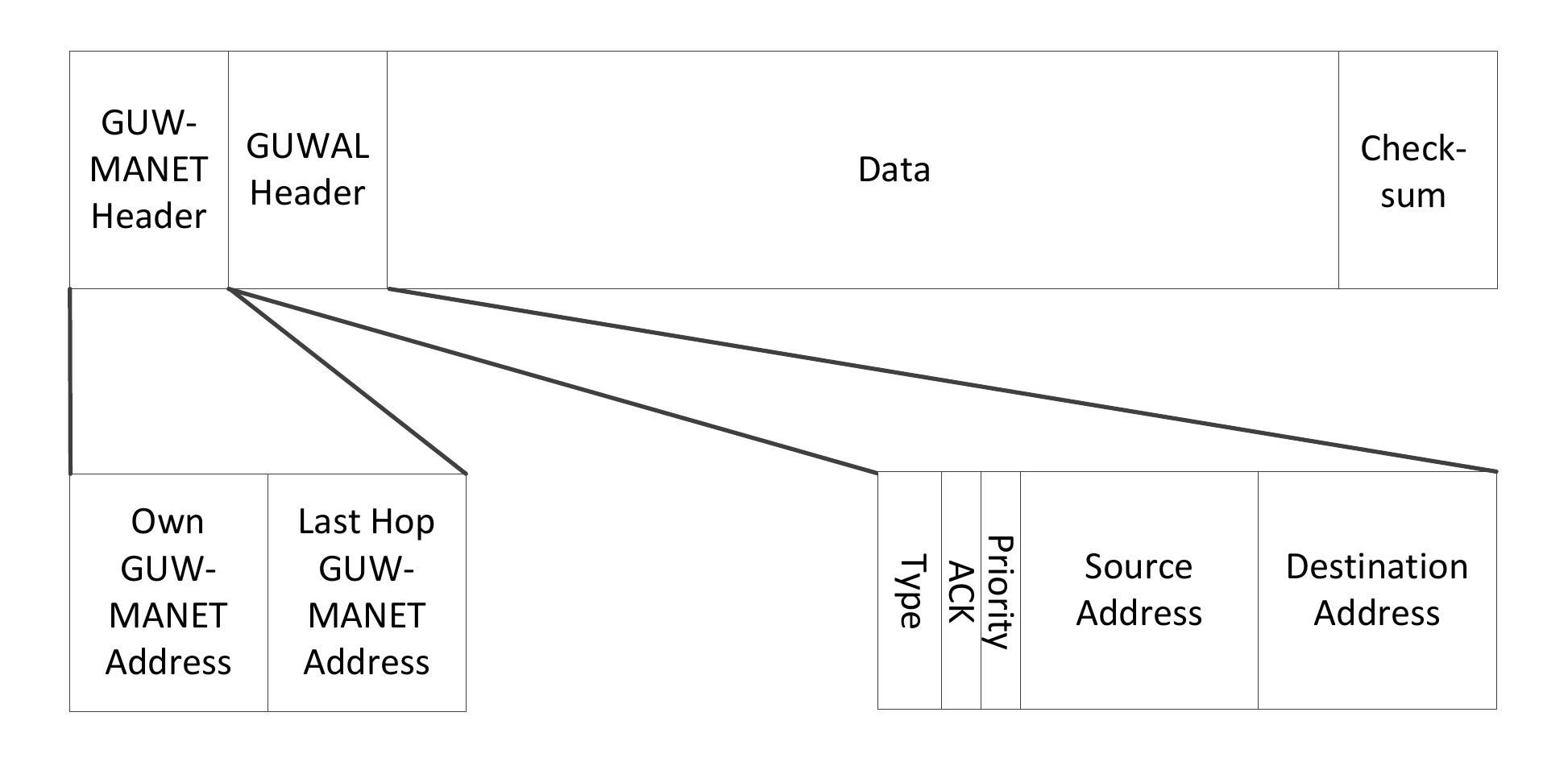}
\caption{The GUWMANET/GUWAL protocol.}
\label{proto_guwal}
\end{figure}
\end{center}

The application layer uses the frame format GUWAL (Generic Underwater Application Language) from FWG/FKIE~\cite{guwal}. 
GUWAL is an operational application language for tactical
messaging in underwater networks with low bandwidth. It is based on 16 byte data frames, 
which include 2 bytes for a header in the beginning and 2 bytes for a CRC checksum at the end.
The structure of the GUWMANET/GUWAL protocol is depicted in Fig.~\ref{proto_guwal} (GUWMANET is described in the following section).

The GUWAL header contains a type field, a priority and an acknowledgement flag,
two short operational addresses with 6 bit each, a source and a destination address. 
By only consuming 6 bit, source and destination addresses are designed smaller than regular 32 bit IPv4 addresses in order to save bandwidth. 
Nevertheless, the address range is sufficient for small acoustic mesh networks.
Additionally, each GUWAL address can also be used as multicast address, if necessary. 

Some potential applications for covert acoustical mesh networks based on GUWAL are described in detail
in Section~\ref{covnet_sec5}. For our experiments we implemented an example application that
parses UTF-8 characters from STDIN. 
After the GUWAL header one additional byte is used at the application layer to specify the type of payload.
Each data frame can include a payload of up to eleven characters.
Frames are sent via a TCP connection to the network layer where they are further processed for output.

\subsection{Network Layer}
In the network layer we use the ad-hoc routing protocol GUWMANET (Gossiping in Underwater Mobile Ad-hoc NETworks) from FKIE/FWG~\cite{guwmanet} as the counterpart
to GUWAL. GUWMANET reuses the two 6 bit operational source and destination addresses 
of GUWAL for network routing. Due to the fact that these addresses can be used as multicast 
addresses they do not have to be unique in the network. GUWMANET introduces
a new 5 bit network address, which should be unique in the local 2 hop neighborhood. This allows
to distinguish between all neighbors, which is important to build up routes. In the current
version of GUWMANET these network addresses must be set statically. It is intended by FWG/FKIE to
implement an automatic network configuration procedure in a future version.

The network header of GUWMANET consists of two address fields, as can be seen in 
Fig.~\ref{proto_guwal}. The first address field contains the network address of the
current transmitter. The second field contains the network address of the previous node 
which forwarded the message before (last hop). While the source and destination addresses
of the GUWAL header are end-to-end addresses and static
during the complete forwarding 
process, the transmitter and last hop address in the GUWMANET header changes at each hop.

The route establishment operates on the principle of reactive routing protocols. The first
message is flooded through the network. Each node repeats the message and sets the last hop
field to the network address of the first node it received the message from. A node generates
a temporary routing entry if it overhears that it was selected as last hop in a forwarding 
from one of its neighbors. After the message reaches one of the destination nodes,
it replies with an acknowledgement that is routed back with the temporary routing entries
to the source. On the way back, the route is persisted for all further packets. If a route 
breaks down, the route establishment is initiated again. 
Further details on the routing protocol are provided by Goetz and Nissen~\cite{guwmanet}.

\subsection{Link Layer}
\subsubsection{General Considerations}
At the link layer, we provide the PHY application, which includes the modem of the Adaptive Communication System (ACS), originally developed by FWG and based on GNU Radio~\cite{Blossom:2004}, an open-source development toolkit for signal processing. 

\subsubsection{Error Correction Layer}
Both modems can benefit from the optional EC (error correction) layer from FWG/FKIE. The EC layer uses the 16 bit checksum included at the end of each packet to restore packets
with one or two bit errors. If a node receives multiple error-prone versions of the message, the EC layer tries to merge them into a correct one.
More details on the error correction will be included in a future article.

\subsubsection{Physical access using ACS modem}

\begin{center}
\begin{figure}[!ht]
\includegraphics[width=0.5\textwidth]{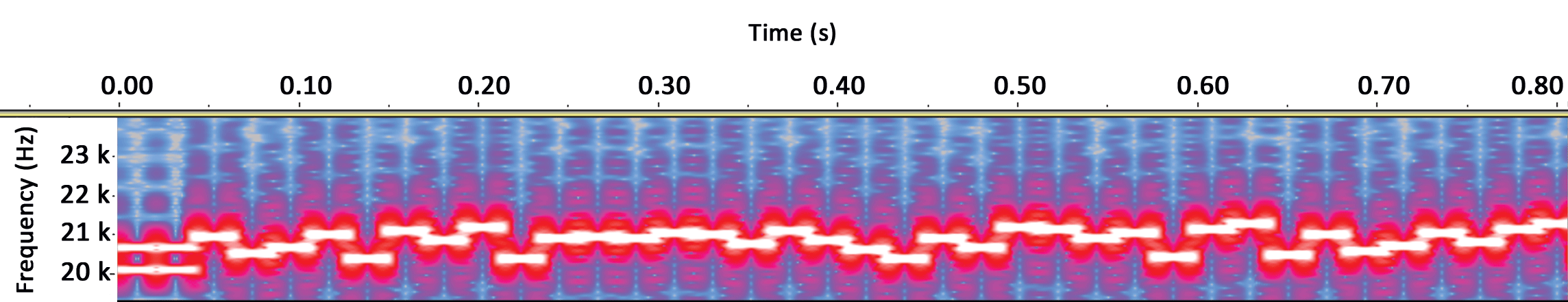}
\caption{Signal spectrogram---The results of the utilized frequency hopping algorithm are made visible.}
\label{spectrogram}
\end{figure}
\end{center}

The ACS modem uses JANUS~\cite{janus}, a robust signaling method for underwater communications.
It is based on FHSS (frequency-hopping spread spectrum) with 48 carriers (20 carriers are used for pure ultrasonic communication in this article).
Fig.~\ref{spectrogram} shows the spectrogram that reveals the frequency hopping within a transmission of the ACS modem.

The center frequency was modified from 4,200 Hz to 21,000 Hz (18,600 Hz with the originally presented waveform~\cite{hanspach.jocm}), completely placing the signal into the ultrasonic frequency range.
In order to avoid undersampling, the sampling frequency was modified from 12,800 Hz to 48,000 Hz and the sample size for each bit value was modified from 256 to 1024.
It could be observed that the ACS modem (running on top of a Lenovo T400) was not capable of calculating the IFFT/FFT of size 1024 fast enough in software.
Therefore, for any bit position $j$ within a output frame with the associated output frequency $\lambda(j)$ within the frequency hopping sequence $\lambda$ and
for each sample position $k$, the sample $s_k$ is calculated by Eq.~(\ref{eq_covnet}).

\begin{equation}\label{eq_covnet}
s_k = \frac{1}{1024} \cdot e^{2 \pi i \cdot \frac{\lambda(j) k}{1024}} \cdot w_k
\end{equation}

To obtain a better differentiation between each bit, we multiply the sample with the window function $w_k$, which implements a trapezoid window and the square root of a Hamming window, which is defined by Eq.~(\ref{eq_covnet2}) for each index $i$.
\begin{equation}\label{eq_covnet2}
\sqrt{0.54 + 0.46\; \cdot \cos \left ( \frac{2\pi \left(i - 1024 / 2 \right)}{1024} \right)}
\end{equation}

\begin{center}
\begin{figure}[!ht]
\includegraphics[width=0.5\textwidth]{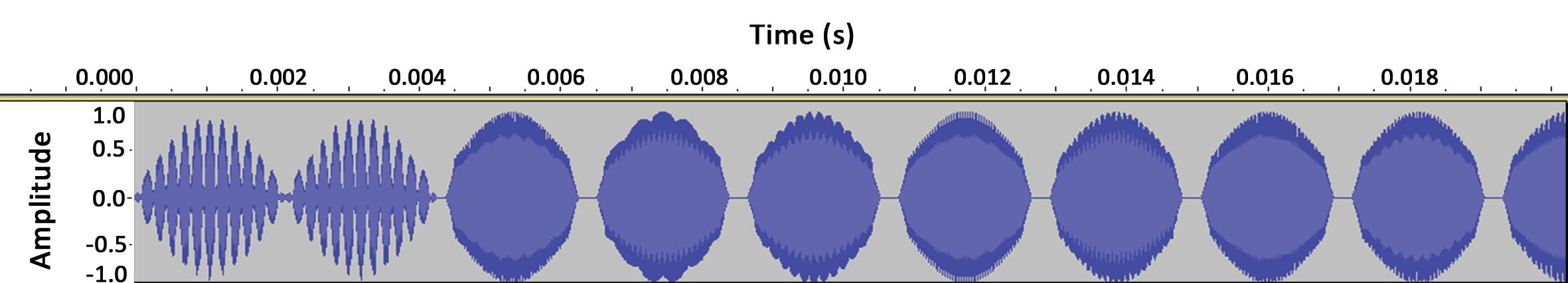}
\caption{Signal waveform---The preamble and 6 transmitted bits are made visible.}
\label{waveform}
\end{figure}
\end{center}

The resulting waveform is depicted in Fig.~\ref{waveform}.

\begin{center}
\begin{figure}[!ht]
\includegraphics[width=0.49\textwidth]{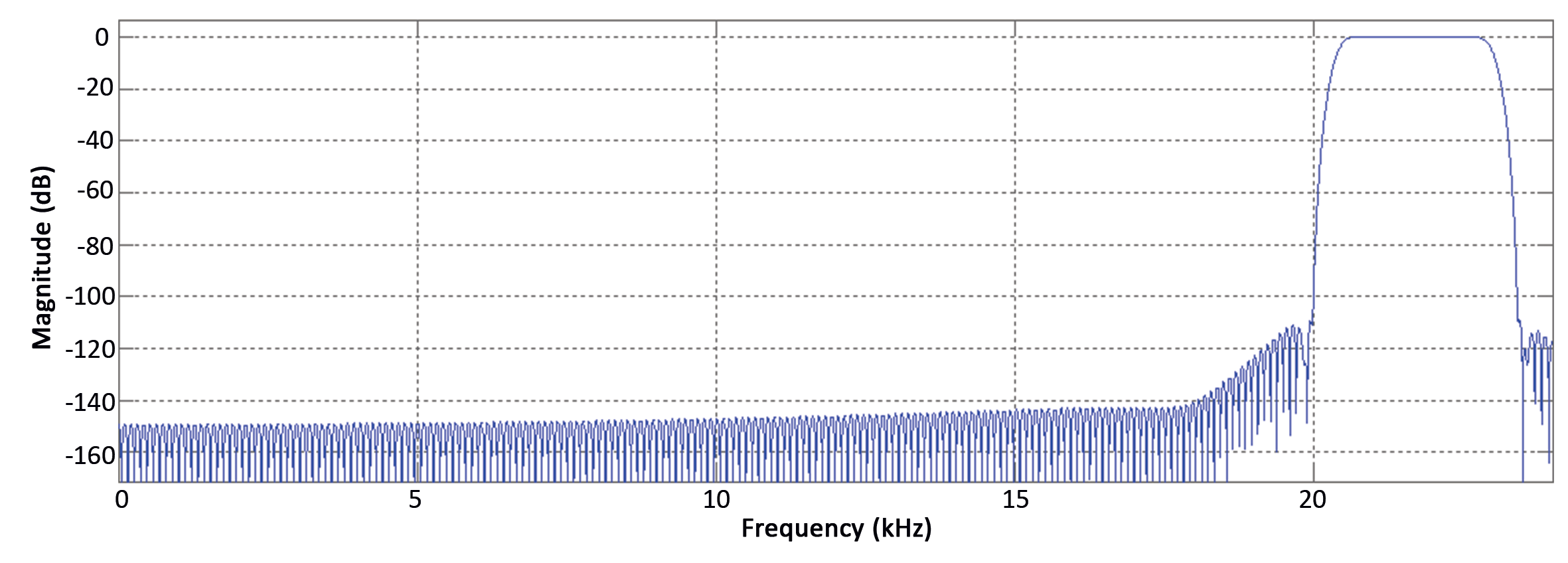}
\caption{The bandpass FIR filter designed (Blackman-Harris window).}
\label{bandpass}
\end{figure}
\end{center}

In the first 42 ms a preamble for packet detection and synchronization is sent. 
We added an adapted bandpass filter at the receiver of the original ACS modem in order to filter for the transmitted frequency range (Fig.~\ref{bandpass}).

The bandpass FIR (finite impulse response) filter is a Blackman-Harris window with a frequency range between 20,400 and 23,000 Hz.

\section{Experiments and Measurements}\label{covnet_sec4}

\subsection{Experiment Setup}

For the experimental setup we are using five laptops (model: Lenovo T400) as the mesh network participants.
As operating system for each node, we installed Debian 7.1 (Wheezy) on each laptop.
All experiments were performed at FKIE, building 3, and without any acoustical preparations made.

\subsection{Output Frequency Measurement}\label{output_fr}

\begin{center}
\begin{figure}[!ht]
\includegraphics[width=0.49\textwidth]{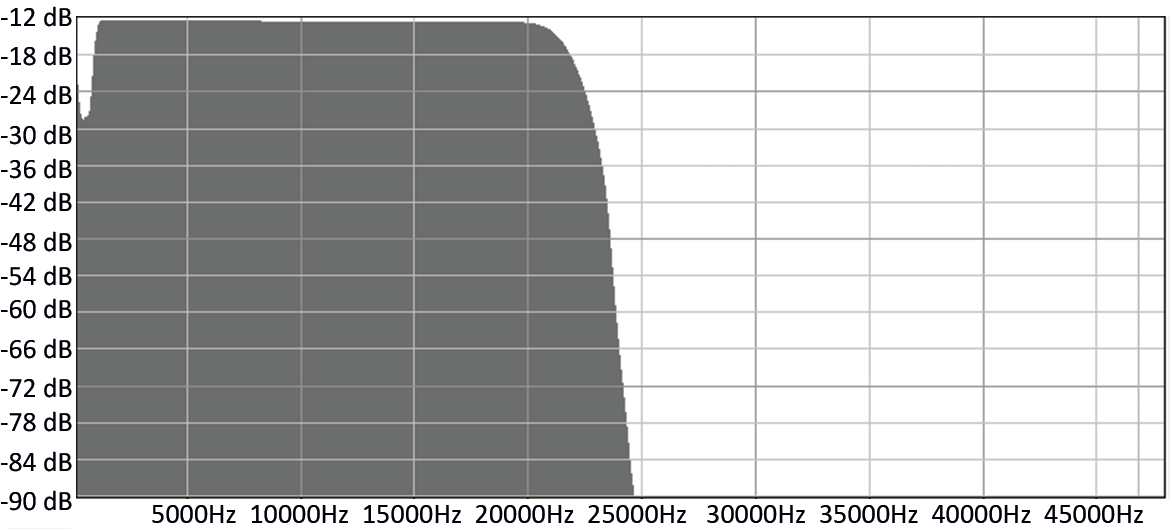}
\caption{The recorded output/input frequency response of a Lenovo T400 Laptop (Hanning window, FFT, size 512).}
\label{freq_t400}
\end{figure}
\end{center}

The Lenovo T400 features an Intel Corporation 82801I (ICH9 Family) HD Audio Controller.
The frequency range that the audio processor is able to output and input is determined by directly connecting the line in and line out jacks and recording an increasing signal from 0 to 35,000 Hz. 
The resulting graph is depicted in Fig.~\ref{freq_t400} where the output sound pressure levels are correlated with the output signal frequency. 
The recorded sound pressure levels are relative values that describe the attenuation of the original audio signal.

From the recorded frequency range it can be discovered that we are able to process frequencies in the low ultrasonic range around 20,000 Hz. 
Previously performed tests with a Lenovo T410 Laptop featuring the Conexant 20585 audio codec (with 192 kHz DAC / 96 kHz ADC)~\cite{conexant} have shown very similar results.
The results lead us to the conclusion that ultrasonic or near-ultrasonic communication with computing systems of the Lenovo T400 series is possible.

\begin{center}
\begin{figure}[!ht]
\includegraphics[width=0.49\textwidth]{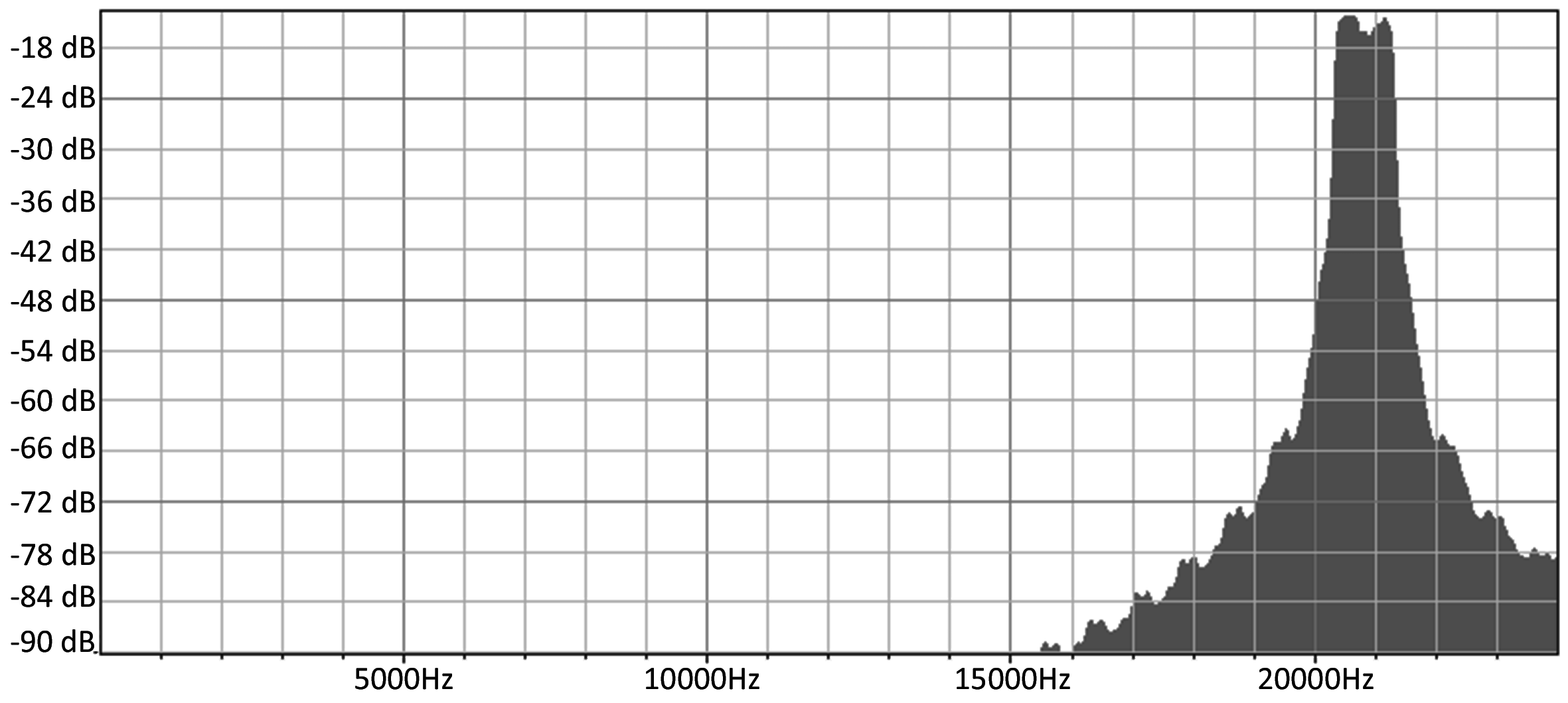}
\caption{The recorded frequency range of an ACS modem transmission (Hanning window, FFT, size 1024).}
\label{freq_acs}
\end{figure}
\end{center}

Fig.~\ref{freq_acs} shows the output frequency spectrum of a pure ultrasonic transmission with the ACS modem.

The original preamble of the ACS modem has been replaced by a 0.042 s long preamble (see also waveform, Fig.~\ref{waveform}), which is also placed in the ultrasonic frequency range.
The transmission was found to be inaudible to the observers of the experiment at the configured volume levels.

\subsection{Range Experiment}

\begin{center}
\begin{figure}[!ht]
\includegraphics[width=0.5\textwidth]{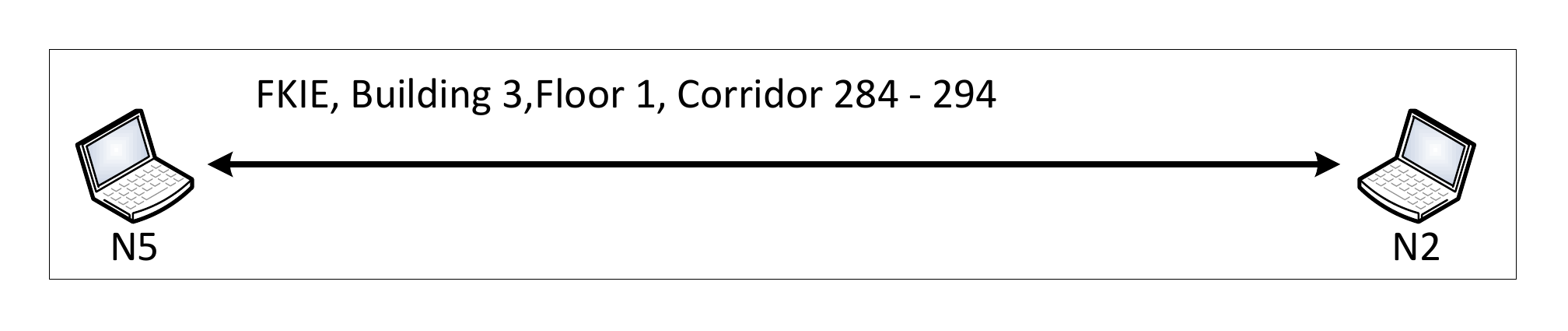}
\caption{Schematic view of the range experiment.}
\label{range_fkie}
\end{figure}
\end{center}

In another experiment we interconnect two laptops in order to gain an understanding of the achievable range in a covert acoustical mesh network.
Acoustic waves are attenuated during transmissions in air (or water) due to scattering, absorption and reflection, leading to lower sound pressure levels in the received signal.
The degree of attenuation is specifically depending on the distance between transmitter and receiver and on the frequencies used in data transmission. 
Therefore, we should be able to communicate over longer distances with sub-ultrasonic frequencies in comparison to ultrasonic frequencies.
Fig.~\ref{range_fkie} shows the setup of the range experiment.

As depicted, the range experiment is performed in an approximately 25 m long corridor at FKIE.
Both nodes are placed in direct line of sight to each other with the displays directed at each other so that output of the internal speakers (built-in alongside the keyboard) is loosely directed at the corresponding node.
The achievable range is determined by repeatedly transmitting messages and gradually increasing the distance between the nodes with each successful transmission.

\subsubsection{Measurements with ACS Modem}

With the ACS modem, transmissions over a distance of up to 8.2 m could be observed with a transmission rate of approximately 20 bit/s.
Moreover, distances of 19.7 m have been overcome with the originally presented waveform that is placed slightly below the ultrasonic frequency range~\cite{hanspach.jocm}).
As it was already presented in Fig.~\ref{freq_t400}, the frequency response of the utilized audio hardware gradually decreases for frequencies $\geq$ 20,000 Hz, and this is one of the reasons that the near-ultrasonic frequency range below 20,000 Hz (which might be considered less stealthy in comparison to the ultrasonic frequency range $\geq$ 20,000 Hz) is more stable in covert acoustical transmissions.
No bit errors could be observed in transmissions over these ranges.
While much higher transmission rates are generally possible with acoustical communication (using a less narrow band), tests with a configured transmission rate of 20 bit/s have been the most successful in transmissions over distances $\geq$ 5 m.
Thus, we conclude that the achievable transmission rate is highly dependent on the desired transmission distance.

\subsection{Interconnection Experiment with ACS Modem}

\begin{center}
\begin{figure}[!ht]
\includegraphics[width=0.5\textwidth]{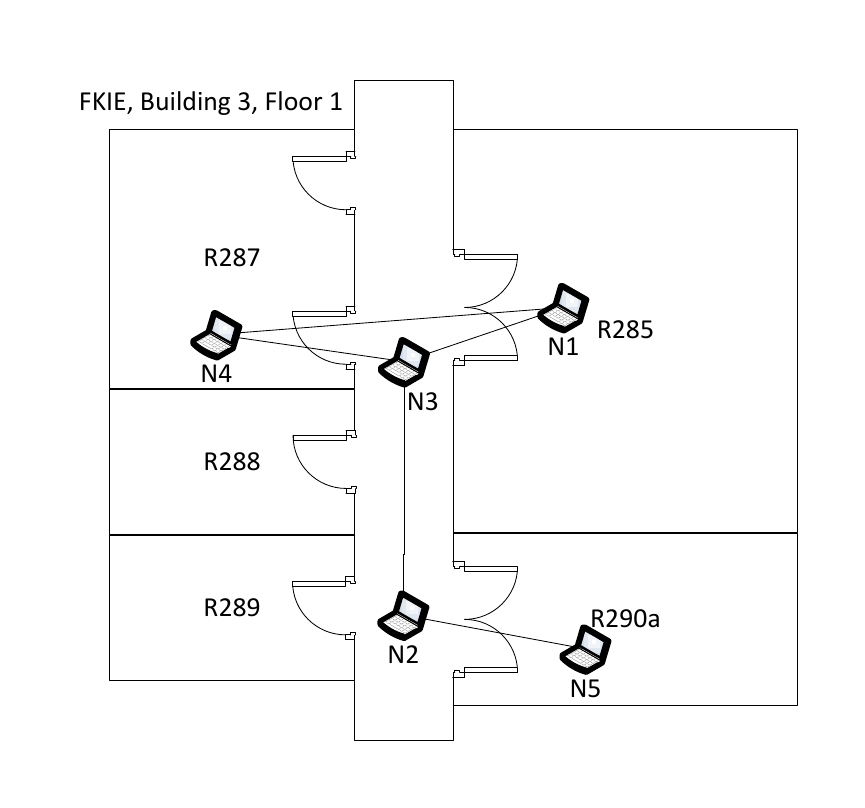}
\caption{Schematic view of the interconnection experiment.}
\label{rooms_fkie}
\end{figure}
\end{center}

Fig.~\ref{rooms_fkie} shows the experiment setup for a covert network of 5 participants with the adapted ACS modem.

The nodes are placed in a mixed office/lab environment at FKIE.
Every node is placed in direct line of sight to another node in order to establish connectivity.
While indirect communication over reflections from walls and doors has also been tested, connectivity could only be established over direct communications.
Tab.~\ref{dist_covnet} shows the distances between connected nodes in the performed experiment.

\begin{table}[!ht]
\centering
\caption{Distances overcome between connected nodes in the interconnection experiment.}\label{dist_covnet}
\renewcommand{\arraystretch}{1.3}
\begin{tabular}{ | c | c | }
\hline
\textbf{Connected Nodes} & \textbf{Distance} \\
\hline
N5 $\leftrightarrow$ N2 & 3.4 m \\
\hline
N2 $\leftrightarrow$ N3 & 5.4 m \\
\hline
N3 $\leftrightarrow$ N4 & 2.8 m \\
\hline
N3 $\leftrightarrow$ N1 & 3.3 m \\
\hline
N4 $\leftrightarrow$ N1 & 6.2 m \\
\hline
\end{tabular}
\end{table}

As one GUWMANET/GUWAL packet is transmitted over a duration of approximately 6 s, a three-hop transmission (using 4 computers) would have a latency of 18 s, which could be confirmed in the experiment setup.
Every sent packet could be delivered successfully (no bit errors at destination) in the 5-nodes-network, either in the first transmission or in one of the three automatic retransmissions.
With the equiripple bandpass filter in place (see Fig.~\ref{bandpass}), common sources of noise (such as human speech) were not found to have a considerable effect on the transmissions.
In another test, the absorption of acoustic waves by humans, walking through the experiment setup and, therefore, blocking the line of sight between two nodes, was found to have an adverse effect on connectivity.
However, this does not affect the operation of a covert acoustical mesh network, as transmissions without acknowledgments are automatically retransmitted.
Now that we have shown the feasibility of implementing covert acoustical mesh networks, we will have a look at potential applications for these covert networks.

\section{Applications of Covert Acoustical Mesh Networks}\label{covnet_sec5} 

\subsection{A Multi-Hop Acoustical Keylogger}

\begin{center}
\begin{figure}[!ht]
\includegraphics[width=0.5\textwidth]{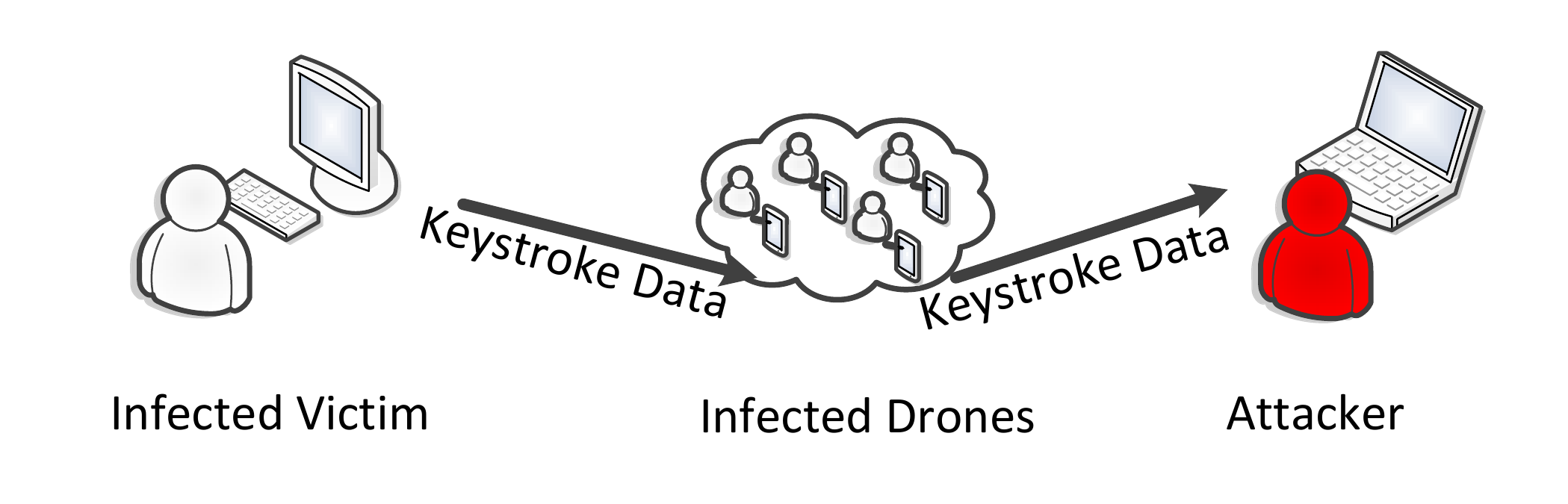}
\caption{Scenario for a multi-hop acoustical keylogger.}
\label{keylog}
\end{figure}
\end{center}

We use the keylogging software logkeys~\cite{logkeys} for our experiment on acoustical multi-hop keylogging.
The considered scenario is depicted in Fig.~\ref{keylog}.

\begin{center}
\begin{figure}[!ht]
\includegraphics[width=0.5\textwidth]{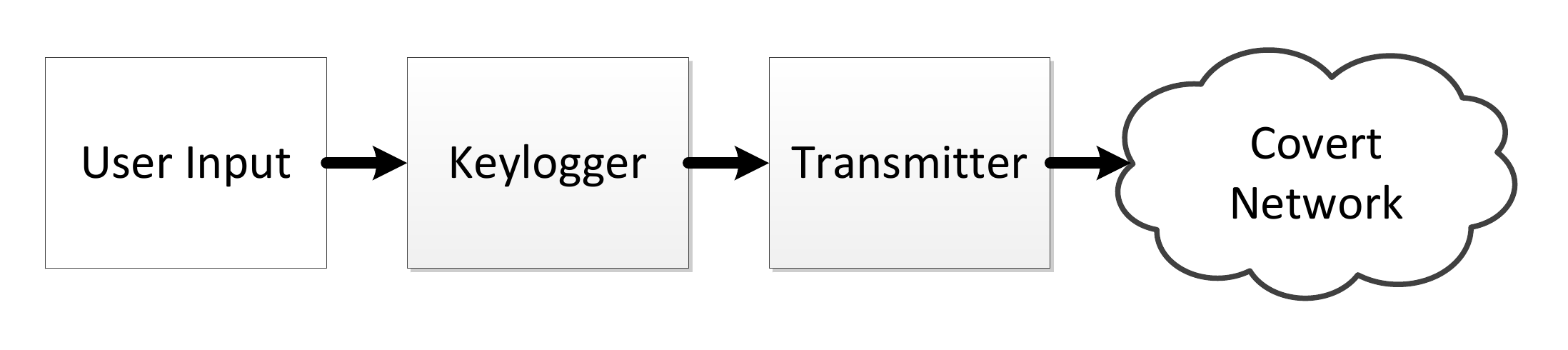}
\caption{A multi-hop acoustical keylogger that is maliciously placed at the infected victim.}
\label{namedpipe}
\end{figure}
\end{center}

The infected victim sends all recorded keystrokes to the covert acoustical mesh network.
Infected drones forward the keystroke information inside the covert network till the attacker is reached who is now able to read the current keyboard input of the infected victim from a distant place.
To implement the described scenario, we propose a multi-hop acoustical keylogger (Fig.~\ref{namedpipe}).

User input is recorded by the keylogger \emph{logkeys} that must be in place and started with root privileges at the infected victim.
The keylogger is configured to write any keystrokes to a named pipe that is read out by the acoustic transmitter.
The transmitter becomes active when a line feed symbol is reached and sends the keystroke information out to the covert network.
The multi-hop acoustical keylogger has been successfully tested in the previously defined experimental setup.

\subsection{Connecting to and Tunneling over the Internet}

\begin{center}
\begin{figure}[!ht]
\includegraphics[width=0.5\textwidth]{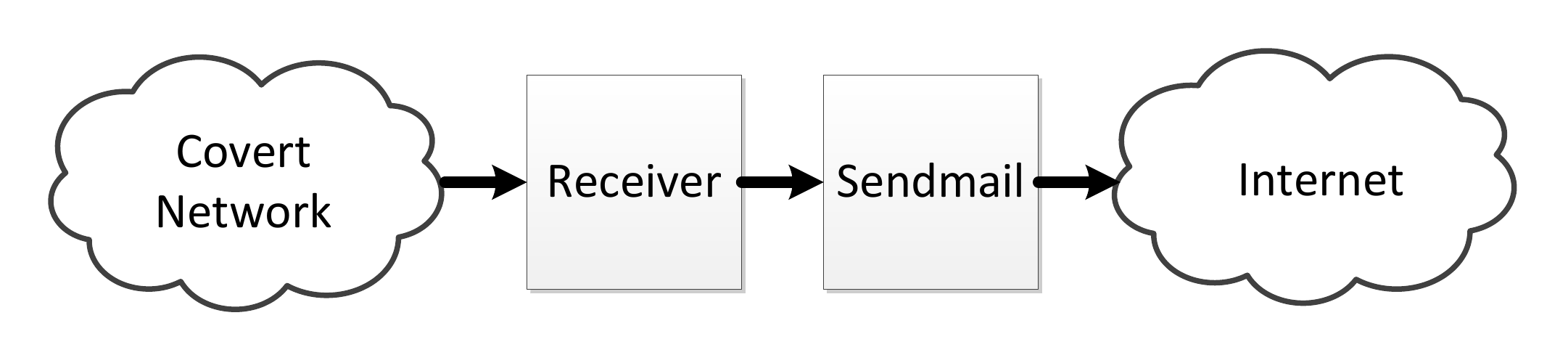}
\caption{An SMTP/TCP/IP proxy that is used by the attacker to forward the keystroke information into the Internet.}
\label{attacker_smtp}
\end{figure}
\end{center}

In a more advanced experimental setup, we connect the attacker to an SMTP server that is connected to the Internet, collecting frames or lines of keystrokes and sending them out as email.
For this purpose, we propose an SMTP/TCP/IP proxy (Fig.~\ref{attacker_smtp}) that encapsulates the data from these frames into the TCP/IP world.

The acoustical receiver gathers input from the covert network and forwards it to a named pipe.
From there, a local mail server (i.e., \emph{sendmail}, but a remote mail server could also be used) is called upon to send an SMTP message out to an arbitrary email address.
This message could just contain the recorded keystrokes, but it is also possible to include the GUWMANET/GUWAL headers in order to tunnel the protocol over HTTP/TCP/IP (circumventing corporate firewalls), and to extend the covert acoustical mesh network to another covert network at any place in the world.
The attacker has to be (directly or indirectly) connected to the Internet in order to perform this procedure, but messages could also be gathered and sent out as soon as Internet access is established.
Sending out gathered keystrokes to a local SMTP server and forwarding them to a remote email server has been successfully performed in the previously defined experimental setup.

\subsection{Other Applications for Covert Networks}

Alongside the presented proof-of-concept, even more applications of covert networks are conceivable.
Besides keystroke information it would also be possible to forward other security critical data such as private encryption keys or small-sized text files with classified information from the infected victim to the covert network.
This data could be sent out periodically to maximize the likelihood of data extraction from the host and it could also be spread to different environments when the computing system is carried around.

Finally, an infected drone might not only serve as a router in the covert network, but it might also offer \emph{covert services} to the network.
A covert service is a network service that could, for instance, provide access to further networks (e.g., the Internet, if the attacker is not connected to the Internet itself).

Against malicious participation in a covert acoustical mesh network, countermeasures might be applied as described in the following section.

\section{Countermeasures against Information Leaks from Covert Networks}\label{acous_counter}

\subsection{Implementing Audio Filtering Options}

If audio input and output devices cannot be switched off, implementation of audio filtering options may be an alternative approach to counter maliciously triggered participation in covert networks.
In Linux-based operating systems, a software-defined audio filter can be implemented with ALSA (Advanced Linux Sound Architecture) in conjunction with the LADSPA (Linux Audio Developer's Simple Plugin API) as mentioned by Phillips~\cite{ladspa}.

The specific frequencies used by the presented acoustic modem could be filtered out with a bandpass or lowpass filter.
We tested a 4-pole lowpass filter with LADSPA and a configured cutoff frequency of 18,000 Hz in the presented experimental setup that effectively prevented inaudible communications in any ALSA-based application.

A general approach for audio filtering in a component-based operating system might be to implement a trusted audio filtering component, i.e., an audio filtering guard.
The audio filtering guard would be connected to the audio input and output devices and to the operating system partitions $p_i$ and $p_j$ that need access to the audio devices.
For all audio input and output operations, the audio guard would apply input and output filters.
To ensure that audio filtering is always-invoked in audio input and output, the access control policy of the component-based operating system needs to enforce that audio input and output devices can only be accessed via the audio guard.
More details on the concept of an audio filtering guard are provided by Hanspach and Keller~\cite{hanspach.passat}.

A more advanced approach might be to implement an audio intrusion detection system (IDS) that does not only filter along predefined settings but supports methods for detection of modulated audio signals and handling input and output based on the signal characteristics (e.g., as recently presented by us~\cite{hanspach.sich14}).
This audio IDS might also be implemented as an operating system guard, as will be described in the following section.

\subsection{A Host-Based Audio IDS Designed as an Operating System Guard}

\begin{center}
\begin{figure}[!ht]
\includegraphics[width=0.5\textwidth]{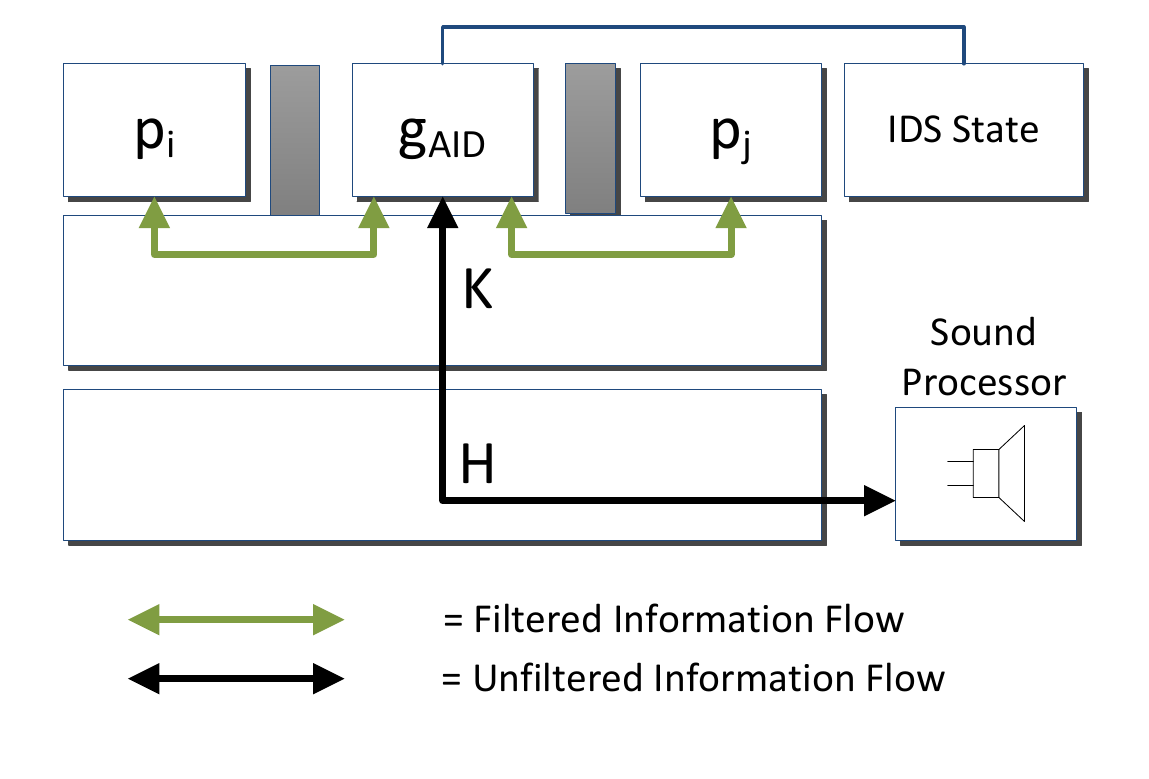}
\caption{Design concept of an audio intrusion detection guard.}
\label{guard_ids}
\end{figure}
\end{center}

The proposed architecture of a host-based audio intrusion detection guard (Fig.~\ref{guard_ids}) is similar to the described audio filtering guard, but an IDS state may be included for stateful inspection of audio input and output.

The audio intrusion detection guard would forward audio input and output signals to their destination and simultaneously store them inside the guard's internal state where they are subject to further analyses.
Signal analyses might include frequency, amplitude and phase measurement in order to detect digital signal processing modes.
A decision routine inside the guard would apply filters to the audio interfaces so that any audio input from and output to operating system partitions is filtered.
For instance, it is possible to pitch highly suspicious ultrasonic signals down in order to inform the user of a hidden audio transmission (we have also demonstrated this with a mobile detection device, see also~\cite{hanspach.sich14}).
Further information on the captured signal would be included in the operation system's log files in order to enable in-depth assessment of a possible attack.

The concept of an intrusion detection guard analyzing device input and output might also be adapted to other types of physical emanations such as modulated optical transmissions that have been described by Frankland~\cite{frankland}, Loughry and Umphress~\cite{Loughry:2002:ILO:545186.545189}, and Hasan, Saxena, Tzipora, Shams and Dustin~\cite{Hasan:2013:SCH:2484313.2484373}.
Considerations on the systematic identification of these covert physical channels will be presented in an upcoming publication~\cite{hanspach.futsec}.
Related studies are further discussed in the following section.

\section{Related Work}\label{covnet_sec7}

As White et al.~\cite{cps} pointed out, mobile cyber-physical system applications might be present on common sensory devices (e.g., on smartphones).
Malware infected devices might be connected to form a botnet (as described by Dagon, Gu and Lee~\cite{botnets}).
With smartphones and similar devices, such a botnet might be built upon one of many network interfaces present in such devices and even on input and output sensors as we describe in this article, although we do not specifically experiment with smartphones and focus at interconnecting laptops.
A near-field communication botnet approach based on bluetooth is presented by Singh, Jain, Traynor and Lee~\cite{singh:evaluating}.
We, instead, use audio communications for our botnet-like approach in near-field communications to construct a covert acoustical mesh network consisting of malware infected drones.

Hasan, Saxena, Tzipora, Shams and Dustin~\cite{Hasan:2013:SCH:2484313.2484373} present a botnet approach where different types of physical emanations are used for command-and-control in botnets.
In contrast to these authors, we are not only looking at command-and-control messages, but we build a whole covert mesh network from acoustical emanations.

Different types of physical emanations have been discussed as means for information leaking.
Van Eck~\cite{vanEck:1985:ERV:7307.7308} describes the risk from electromagnetic radiations, while Loughry and Umphress~\cite{Loughry:2002:ILO:545186.545189} present information leakage from optical emanations.
Shamir and Tromer~\cite{cryptana}, and LeMay and Tan~\cite{lemay} describe acoustical emanations from the computing system (e.g., supply capacitors from the computer's mainboard) that might leak critical information.
We are also using acoustical emanations for information leakage, but instead of passively monitoring the computing system in order to exploit a side channel, we actively produce acoustical emanations to construct a covert channel between different computing systems and interconnect them into a covert acoustical mesh network.

For eavesdropping on a computer user's keystrokes by using a computers's physical emanations, several techniques have been proposed.
Halevi and Saxena~\cite{Halevi:2012:CLK:2414456.2414509} present a study on acoustical emanations from keyboards where the keyboard input is recovered from the sound of a keystroke.
Raguram, White, Goswami, Monrose and Frahm~\cite{Raguram:2011:IAR:2046707.2046769} discuss a setup where typed input is recovered from environmental reflections, while Frankland~\cite{frankland} describes information leakage over keyboard LEDs that are manipulated to carry an optical signal.
Moreover, Balzarotti, Cova and Vigna~\cite{10.1109/SP.2008.28} show that keystrokes can be automatically captured by recording the keyboard with a camera and analyzing the video stream.
We also gather keystrokes from a computer user, but we record the keystrokes directly on the infected victim (i.e., the user's computer) where a malware prototype needs to be placed in preparation of an attack.
This malware prototype contains both a keylogger software and a network stack for covert acoustic communication in order to spread the recorded keystrokes over the covert network.

Using audio as a networking technology has been described by Madhavapeddy, Scott, Tse and Sharp~\cite{Madhavapeddy:2005:ANF:1083818.1083942}, by Yan, Zhou, Shi and Li~\cite{Yan:2007:DIO:1287812.1287831} and by Lopes and Aguiar~\cite{10.1109/MPRV.2003.1228528}.
Very recently, Nandakumar, Chintalapudi, Padmanabhan and Venkatesan~\cite{Nandakumar:2013:DSP:2486001.2486037} have presented a study on acoustic peer-to-peer near-field-communication (up to 15-20 cm), where eavesdropping by third parties is prevented by the means of signal jamming.
Based on the concept of audio networking, we construct a wireless mesh network, not with the ambition to compete with wireless communication standards, but to prove that covert and stealthy communication is possible with audio networking and that critical information can be leaked over a covert acoustical mesh network.

Furthermore, acoustic wave propagation is regularly used in underwater setups as described by Otnes et al.~\cite{otnes} and many more authors. 
We, however, are using a setup for covert air communications while utilizing parts of an underwater communication system~\cite{wuwnetEmulator}.

Preliminary work on ultrasonic communication has been performed by Hosman, Yeary, Antonio and Hobbs~\cite{5488066}, by Tofsted, {O'Brien}, {D'Arcy}, Creegan and Elliott~\cite{arl}, by Altman, Antebi, Atsmon, Cohen and Lev~\cite{ultra}, and by Li, Hutchins and Green~\cite{4494786}.
In contrast to these authors, we show how ultrasonic communication can be placed in the context of information security and how security critical data can be leaked over multiple hops of infected drones.
 
On behalf of the UCAC Consortium, Leus and van Walree~\cite{Leus:2008:MOC:2312278.2316274} present a setup for stealthy communication with unmanned underwater vehicles where audio messages are transmitted ``at a low signal-to-noise ratio (SNR), thereby hiding messages in the ambient noise and reducing the probability of detection by third parties''.
We also propose a stealthy audio-based communication system, but we primarily aim at providing stealthiness in audio transmissions over the air and from human computer users that might not anticipate information leaks from audio subsystems.
With this scenario in mind, we build a mesh network upon audio transmissions that are inaudible to human computer users by utilizing the ultrasonic frequency range.

\section{Conclusion}\label{covnet_sec8}

We have shown that the establishment of covert acoustical mesh networks in air is feasible in setups with commonly available business laptops.
By reutilizing an underwater communication system, we take advantage of a network stack that was built with robust acoustical communication in mind.
The presented approach to covert acoustical mesh networks allows to transmit messages with a rate of approximately 20 bit/s up to a range of 19.7 m between two connected nodes, but much higher transmission rates would be possible for low-distance transmissions.
The complete path of a single frame from the infected victim at the first hop to the attacker at the last hop over two additional infected drones as intermediate hops took 18 s.

Acoustical networking as a covert communication technology is a considerable threat to computer security and might even break the security goals of high-assurance computing systems that did not consider acoustical networking in their security concept.

We did not specifically address the problem of how to infect a computing system with the malware, but this problem exists with any covert channel technology.

For the prevention of participation in a covert acoustical network, it might not always be acceptable to switch off the audio input and output devices, as they might be needed for IP phone communications and other audio applications.
For these cases it is possible to prevent inaudible communication of audio input and output devices by application of a software-defined lowpass filter.
An audio filtering guard can be used to control any audio-based information flow in a component-based operating system.
Furthermore, a host-based audio intrusion detection system can be utilized, as we already presented in a different article.
The audio IDS analyzes audio input and output to detect modulated signals or hidden messages in audio playback, and it might also be implemented as an operating system guard.

\section*{Acknowledgment}
We would like to thank J{\"o}rg Keller, Jochen F. Giese, Henning Rogge and Tobias Ginzler for their helpful comments.
We also would like to acknowledge Ivor Nissen at the Research Department for Underwater Acoustics and Marine Geophysics (FWG) of the Bundeswehr Technical Center WTD 71, Kiel, Germany, for providing us with access to the source code of an underwater networking software (based on the open JANUS waveform protocol) that we adapted for setting up a covert acoustical mesh network in air. 

\bibliographystyle{IEEEtran}
\bibliography{covert_networks}

\end{document}